\documentclass[10pt,article,twocolumn]{IEEEtran}

\usepackage{balance}

\usepackage{cite}
\usepackage{savesym}
\usepackage{amsmath}
\savesymbol{iint}
\restoresymbol{TXF}{iint}
\usepackage{amsfonts,amssymb}
\usepackage{amsmath,bm}
\usepackage{setspace}
\usepackage{url}
\usepackage{graphicx}
\usepackage[small]{caption2}
\usepackage{fmtcount}
\usepackage{wasysym}
\usepackage[small]{subfigure}
\usepackage{textcomp}
\usepackage{eso-pic}

\usepackage{wireless-utils}

\allowdisplaybreaks
\newcommand{\tabincell}[2]{\begin{tabular}{@{}#1@{}}#2\end{tabular}}

\begin{document}

\title{An Open-Source Research Platform for Embedded Visible Light Networking}

\author{
  \IEEEauthorblockN{Qing Wang$^{\amalg\ddagger}$
	      \quad Domenico Giustiniano$^{\amalg}$
	      \quad Daniele Puccinelli$^{\S}$ } \\
  \IEEEauthorblockA{$^{\amalg}$IMDEA Networks Institute, Madrid, Spain} \\
  \IEEEauthorblockA{$^{\ddagger}$University Carlos III of Madrid, Madrid, Spain} \\
  \IEEEauthorblockA{$^{\S}$University of Applied Sciences of Southern Switzerland, Manno, Switzerland} \\
  \IEEEauthorblockA{\textit{Email: \{qing.wang, domenico.giustiniano\}@imdea.org, daniele.puccinelli@supsi.ch}}
}

\maketitle

\begin{abstract}
Despite the growing interest in Visible Light Communication (VLC), a reference networking platform based on commercial off-the-shelf components is not available yet. An open-source platform would lower the barriers to entry to VLC network research and help the VLC community gain momentum.
We introduce OpenVLC, an open-source VLC research platform based on software-defined implementation. Built around a credit-card-sized embedded Linux platform with a simple opto-electronic transceiver front-end, OpenVLC offers a basic physical layer, a set of essential medium access primitives, as well as interoperability with Internet protocols. We investigate the performance of OpenVLC and show examples of how it can be used along with standard network diagnostics tools. Our software-defined implementation can currently reach throughput in the order of the basic rate of IEEE 802.15.7 standard.
We discuss several techniques that researchers and engineers could introduce to improve the performance of OpenVLC and envision several directions that can benefit from OpenVLC by adopting it as a reference platform.
\end{abstract}

\begin{IEEEkeywords}
Networked VLC, open-source platform, low-cost, software-defined implementation, Linux driver
\end{IEEEkeywords}

\section{Introduction}

The formidable uptake of mobile smart devices is driving an ever increasing demand for wireless data, contributing to the wireless spectrum crunch. As a spectrum-rich alternative to Radio Frequency (RF), Visible Light Communications (VLC) is attracting the interests of both researchers and engineers. VLC also represents an appealing alternative to RF for networked embedded devices, for instance in the internet of things, wearable computing, indoor localization and vehicular networks~\cite{zhou,Epsilon,Liu:2011}. In addition, the adoption of VLC would reduce the health hazards caused by overexposure to RF. 

VLC experimental research in networked embedded systems (Networked VLC) has yet to gain momentum due to the lack of a low-cost reference platform. The drawbacks of the existing experimental work on VLC platforms include its lack of openness, its failure to provide broad support for common networking protocols, and its focus on high-end platforms~\cite{Elgala-2011,Grobe2013,Cossu2012}.

Similarly to how the introduction of the Berkeley motes spearheaded networked embedded systems research a decade ago, we believe that a general-purpose, low-cost, open VLC platform would pave the way to novel networking research directions. This paper takes an initial step toward the adoption of VLC in networked embedded systems and introduces OpenVLC, an open-source software-defined networking platform for fast prototyping. OpenVLC runs on a cost-effective yet powerful embedded board, with a unit cost of approximately sixty dollars. The source code and electronic schematic of OpenVLC are available at the following URL: {\url{http://openvlc.org}}.

In this article, we present the design and evaluation of the open-source OpenVLC research platform. We interface an LED-based front-end to an embedded Linux platform and provide a set of software-based primitives, such as signal sampling, symbol detection, coding/decoding, carrier sensing, and communication with the TCP/IP layers of the Linux operating system. We further design and implement a basic Medium Access Protocol (MAC) protocol running in software and illustrate its performance evaluation. The objective of this first step toward Networked VLC is to provide a functional research platform that can be easily extended according to the directions of interest. 

In its present form, OpenVLC relies on simple off-the-shelf electronic components and only uses a basic Physical Layer (PHY), which can be scaled to use more advanced PHYs. Currently, OpenVLC can achieve a MAC layer throughput in the order of the basic rate of IEEE 802.15.7~\cite{IEEE802.15.7}, and UDP throughput of $12.5$~kb/s, {operating  at distances up to $1$~m.}

The rest of this paper is organized as follows. The system design and implementation of OpenVLC are presented in Sec.~\ref{sec_system}, followed by the evaluation at MAC layer and at system level given in Sec.~\ref{sec_evaluation}. Techniques that could improve the performance of OpenVLC and several research directions that can benefit from OpenVLC are discussed in Sec.~\ref{sec_discussion}. Closing remarks are presented in Sec.~\ref{sec_conclusion}.

\section{OpenVLC System Design} \label{sec_system}

\begin{figure}[t]
  \centering
  \subfigure 
	{\includegraphics[width=0.55\columnwidth]{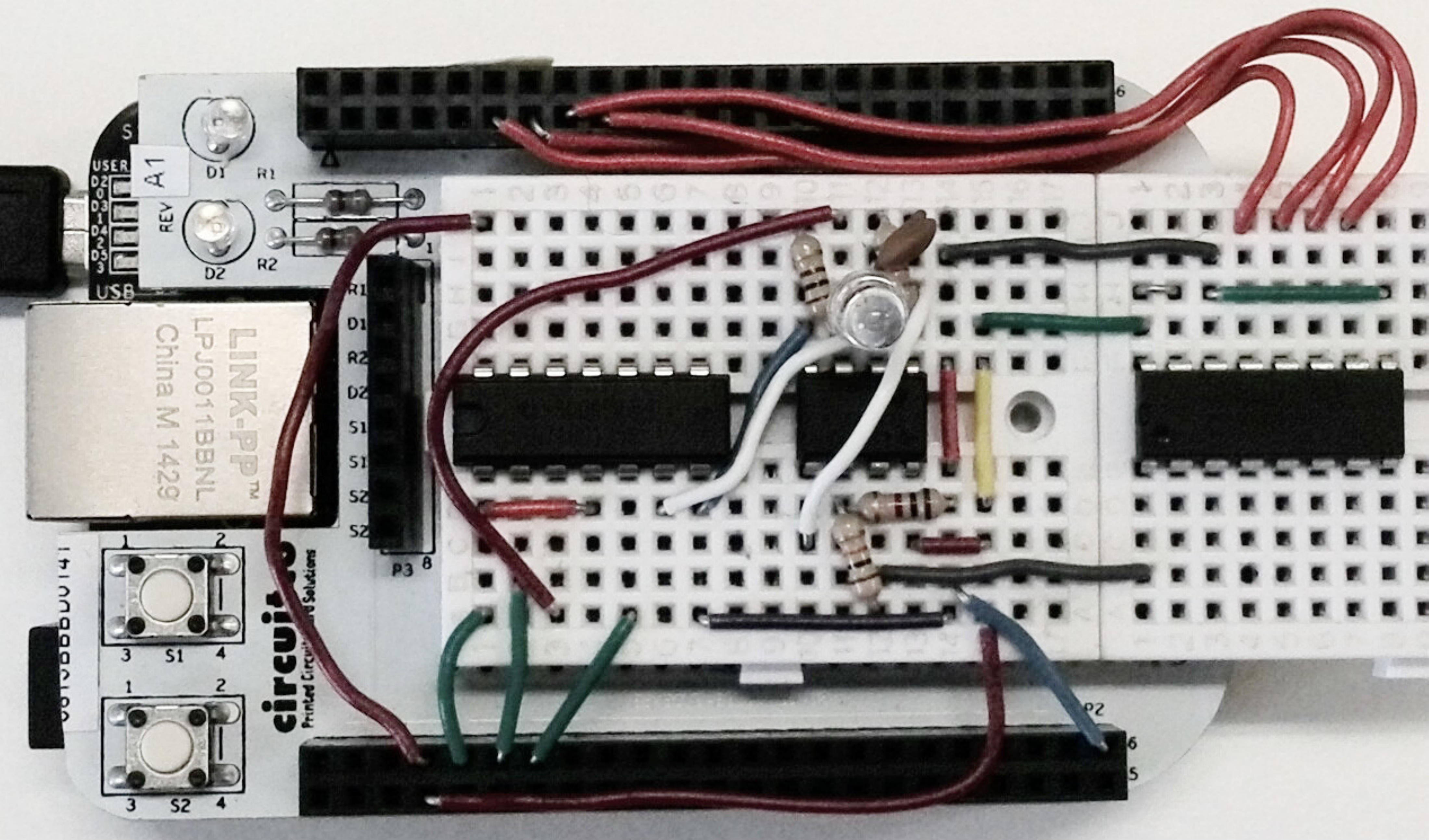}}
  \subfigure 
	{\includegraphics[width=0.42\columnwidth]{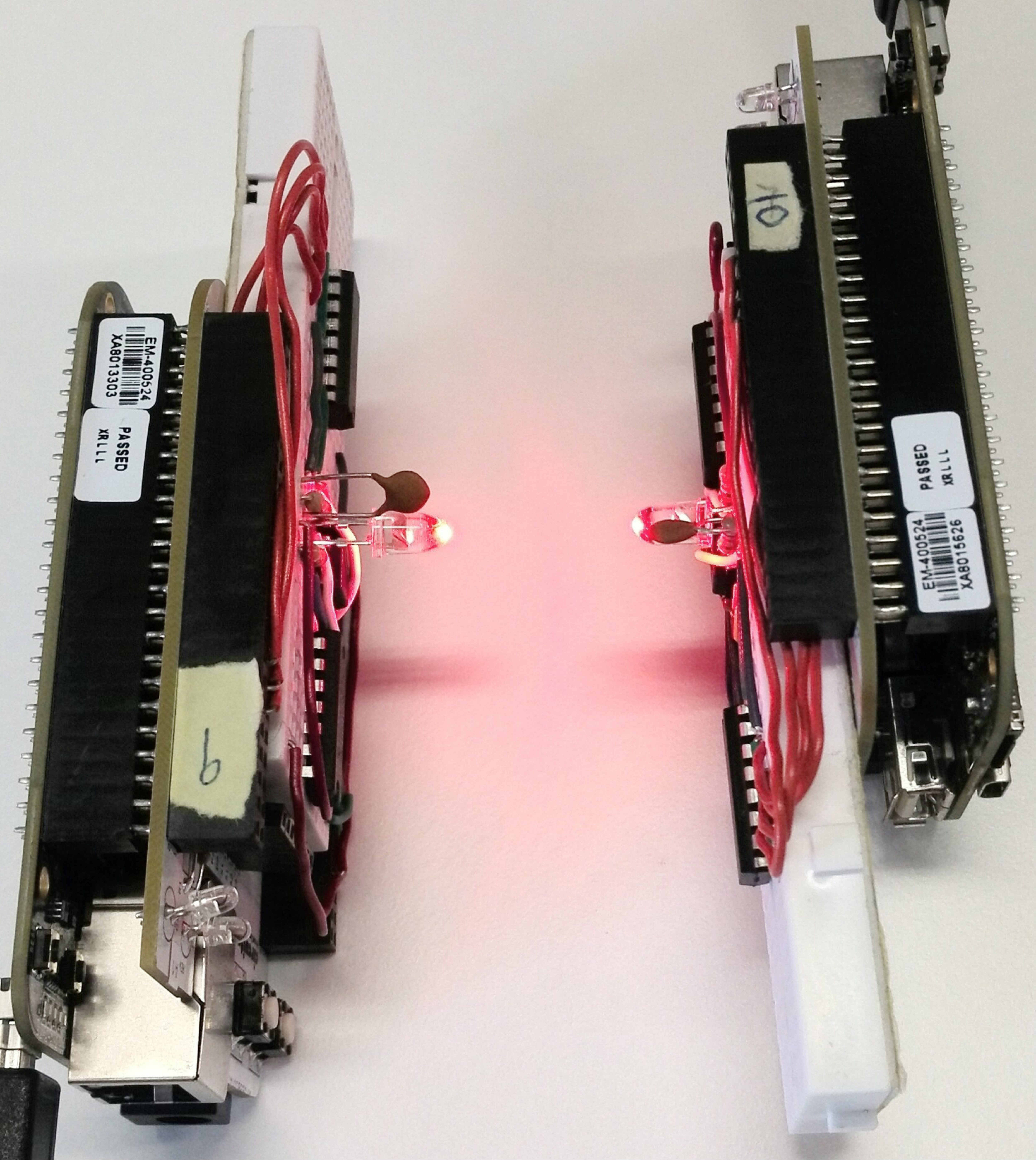}}
  \caption{The current prototype of OpenVLC: the front-end transceiver of an OpenVLC node is shown on the left, and an example of inter-node communication is shown on the right.}
  \label{fig_prototype}
\end{figure}

OpenVLC is a general-purpose software-defined platform for networked VLC. The prototype of OpenVLC is shown in Fig.~\ref{fig_prototype}. It is built around the BeagleBone Black (BBB) board\footnote{\url{http://beagleboard.org/Products/BeagleBone+Black}}, a cost-effective, user-friendly, versatile single-board computer with a small form factor. OpenVLC consists of a BBB board, a VLC front-end transceiver and a software-defined system implementation. The front-end transceiver adopts a single LED together with a few basic electronic components for both transmission and reception. OpenVLC's software components are implemented as a Linux driver that communicates directly with the LED front-end and the Linux networking stack. As a result of this design choice, the VLC communication interface can take advantage of the vast range of Linux tools. The communication between two OpenVLC nodes is illustrated in Fig.~\ref{fig_prototype}.

\subsection{Bidirectional Communication} 

The current version of OpenVLC front-end transceiver reuses the same LED for both transmitting and receiving light signals. Using LEDs as receivers can reduce the design complexity and increases the resilience to ambient noise (e.g., sunlight and indoor illumination~\cite{giustiniano_wd}) with no need for additional optical filters~\cite{Chun2014}. The current design can be extended to use photodiodes as receivers, as we will discuss in Sec.~\ref{sec_discussion}.

\begin{figure*}[t!]
  \centering
  \includegraphics[width=2\columnwidth]{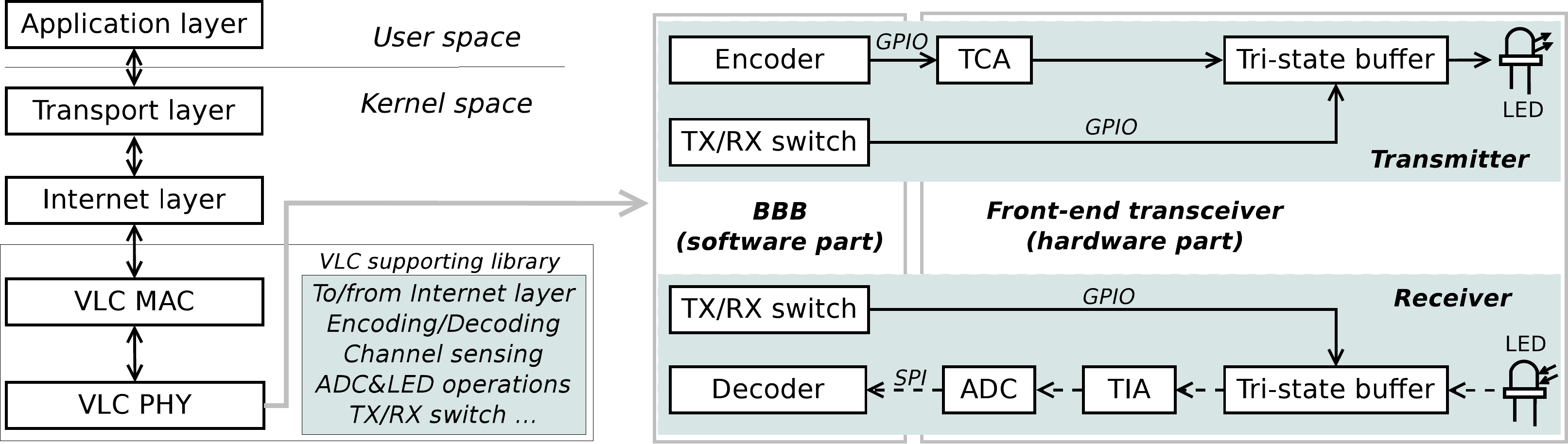}
  \caption{Diagram of the front-end transceiver (right) and the {communication stack} of OpenVLC in an embedded Linux operating system (left).}
  \label{fig_openvlc_structure}
\end{figure*}

The block diagram of the transceiver is shown in the right part of Fig.~\ref{fig_openvlc_structure}. It includes a TransConductance Amplifier (TCA) for transmission, a TransImpedance Amplifier (TIA) and an Analog-to-Digital Converter (ADC) for reception, a tristate-output buffer and ancillary circuitry for transmission and reception. A software-defined Transmitter (TX)/Receiver (RX) switch is used to change the LED operation mode between TX and RX through the GPIO pins:
\begin{itemize}
\item In {\bf TX mode}, the tristate buffer is enabled and encoded signals are first amplified by the TCA and then fed to the forward-biased LED.
\item In {\bf RX mode}, the tristate buffer is disabled to avoid current leakages to the TCA circuitry, and the light signal is received by the reverse-biased LED. The small photocurrent is then amplified by the TIA. Finally, an ADC converts the output analog signals to digital signals, which are then sent to the decoder through the Serial Peripheral Interface (SPI).
\end{itemize}

Through the TX/RX switch and the tristate buffer, OpenVLC can switch the LED between being TX mode and RX mode with low latency, such that it can reliably sustain the operation mode for one or more symbol periods. This design offers a basic setup to implement bidirectional communication using a single LED for VLC networks.

\subsection{Software-Defined PHY Layer} \label{sec_sda}

The communication stack of OpenVLC is illustrated in the left part of Fig.~\ref{fig_openvlc_structure}. Primitives are implemented to build various PHY and MAC layer protocols in the Linux operating system.

\textbf{TX, RX and TX/RX switching.} \label{sec_tx_rx_switching}
In TX mode, the BBB outputs the signal to the anode of LED for a symbol period.  In RX mode, the small photo-current is amplified by the TIA and then sampled by the ADC and converted into a digital signal. The BBB samples the output of ADC at a fixed interval equal to one symbol period. {Symbol boundaries are obtained via the real-time timer of the Linux kernel and handled by our driver. When the timer expires, the TX outputs the signal of the symbol waiting to be transmitted and hold the signal for a symbol period. In turn, the RX samples the output of ADC and stores the value in a sequence that will be decoded by the driver at a later time.} The LED switches between TX and RX mode through the software-defined TX/RX switch that runs on the BBB. 

\textbf{Modulation and detection.}
We adopt intensity modulation for data transmission. Binary information is mapped to the presence (symbol HIGH) or absence (symbol LOW) of the visible light carrier. At the transmitter, we use the On-Off Keying (OOK) modulation and the Manchester Run-Length Limited (RLL) code. Therefore, bit 1 is mapped to symbol sequence LOW--HIGH, and bit 0 is mapped to HIGH--LOW. At the receiver, demodulation is performed with direct detection. Based on the measured voltage, the receiver detects a received signal as a sequence of symbols HIGH and LOW that are then converted to binary data.

\textbf{Preamble.}
The PHY layer transmits each frame with a fixed-length preamble, consisting of an alternate sequence of HIGH and LOW starting with a HIGH symbol. The numbers of HIGH and LOW symbols in the preamble are the same. To convert symbols into binary data, an adaptive symbol detection threshold is adopted because the received light intensity is greatly affected by the free path loss attenuation of light transmitted from the TX to the RX. This detection threshold is obtained on a per-frame basis by averaging out the digital samples of the preamble sequence.
A Special Frame Delimiter (SFD) field is {appended} to the end of the preamble.

\subsection{Software-Defined MAC Layer}\label{sec_mac}
\begin{figure}[b!]
\centering 
  \subfigure{
	\resizebox{\columnwidth}{!}{
	\begin{tabular}{|c|c|c|c|c|c|c|c|c|c|}	\hline
	Preamble & SFD & Length & Dst & Src & Protocol & Payload & CRC \\  \hline
	3B & 1B& {2B} & 2B & 2B & 2B & {0--MAX B} & {2B} \\ \hline
	\end{tabular}} }
  \subfigure {
	\includegraphics[width=\columnwidth]{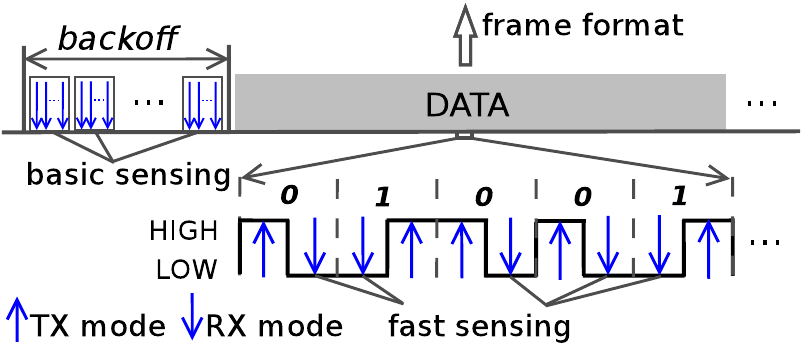}}
  \caption{Backoff, basic and fast sensing of the CSMA/CD protocol. The transmitter uses OOK with Manchester coding to send data. In the frame format: Length$>0\Longleftrightarrow$DATA; Length$=0\Longleftrightarrow$ACK.}  \label{fig_openvlc_mac}
\end{figure}

We define two types of MAC frame: DATA and Acknowledgement (ACK). The frame format is shown in Fig.~\ref{fig_openvlc_mac}. If the frame has no payload (Length=0), it is inferred to be an ACK. Otherwise, it is a DATA frame. 
{Each frame can carry a payload from $0$ to MAX (a predefined value)~bytes.}
The destination and source addresses follow the Length field and each occupies $2$~bytes. The $2$-byte field Protocol identifies the upper layer protocol encapsulated in the frame payload. Fields from the Length to the Protocol form the MAC header. {A two-byte Cyclic Redundancy Check (CRC) over the MAC header and payload is appended after the payload. The Reed-Solomon (RS) error correcting code over the MAC header, payload, and CRC is appended to the end of each frame.}

\textbf{Carrier sensing.}
Wireless MAC protocols usually employ carrier sensing to reduce collisions. In our VLC platform, we provide two types of carrier sensing: \emph{basic sensing} and \emph{fast sensing}. Both are implemented in the PHY layer and can be invoked by the MAC layer. In basic sensing, the platform reads a certain number of continuous symbols. The channel is assessed to be busy if one or more symbols are detected as HIGH symbols; otherwise it is assessed to be clear.
Unlike basic sensing, fast sensing operates on per-symbol basis. The channel is assessed to be clear if the symbol is detected as LOW and is assessed to be busy otherwise. 

\textbf{MAC access protocol.} We implement a MAC layer protocol based on the primitives discussed above.  We employ a contention-based Carrier Sensing Multiple Access/Collision Detection (CSMA/CD) MAC protocol to ensure fair channel access among all VLC nodes and reduce the impact of collisions~\cite{giustiniano_wd}. When a frame is ready for transmission, the MAC first calls the \emph{basic sensing} block of the PHY layer. The frame is transmitted immediately if the PHY layer reports the channel is clear. If the channel is assessed to be busy, the MAC starts a backoff counter. The counter is initialized with an integer value randomly drawn from a uniform distribution within the range (0, CW-1]. The contention window CW is initialized as CW{min}, where CW{min} is the smallest size of the contention window. The PHY layer keeps sensing the channel and each time the channel is assessed to be clear, the counter is decremented. The frame is transmitted when the counter reaches zero.

Upon frame transmission, the transmitter can engage in \emph{fast sensing}. This occurs when the transmitter sends a LOW symbol of the Manchester code, as it powers down the LED and is therefore able to switch the LED to RX mode to receive a symbol, as presented in Sec.~\ref{sec_tx_rx_switching}.
The received symbol is sufficient for \emph{fast sensing}. Afterwards, the LED is switched back to TX mode to carry on the transmission. 
The transmitter alternates between TX and RX mode during data transmission. If the transmitter detects a collision, i.e., the channel is assessed to be busy through \emph{fast sensing} for no less than a predefined interval, the ongoing transmission is immediately interrupted.
The illustration of the backoff mechanism, basic sensing, and fast sensing in CSMA/CD is shown in Fig.~\ref{fig_openvlc_mac}.

After successfully receiving a frame, the receiver sends an ACK to the transmitter. If the transmitter has not received an ACK within the timeout, it retransmits the frame and doubles the CW (until it reaches a pre-defined CW{max} threshold that denotes the maximal size of the contention window).
The frame is dropped after a pre-defined number of failed retransmissions.

\textbf{Interfacing with the Internet layer.}
We implement the MAC protocol as well as part of the PHY layer as a new driver of the Linux operating system. The MAC protocol will become transparent to various applications if it can connect with the Internet layer. 
We implement two primitive functions to receive a packet from the upper layer and the PHY layer, respectively. The first function is called by the Internet layer to move packets to the MAC layer, where they are enqueued for transmission scheduling. The second one receives packets from the PHY layer, checks their protocols, and decides whether or not to send them to the Internet layer. By invoking these two functions, any MAC protocol can easily interact with the Internet layer.

\begin{figure}[t!]
  \centering
  \includegraphics[width=0.96\columnwidth]{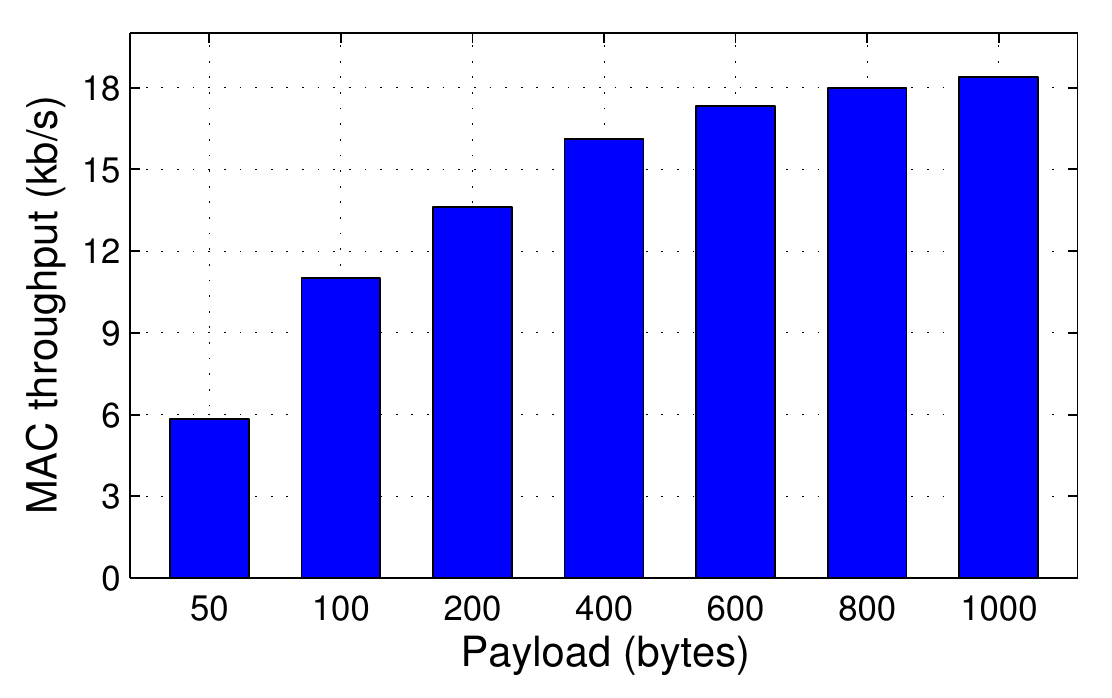}
  \caption {MAC layer throughput as a function of the per-frame payload.}
  \label{fig_mac_throughput}
\end{figure}

\section{Experimental Evaluation} \label{sec_evaluation}
The experimental evaluation described in this section uses the Debian Linux Distribution with kernel version $3.8.13$ and the Xenomai patch. The details of the electronic devices employed in the current version of OpenVLC can be found in~\cite{Wang2014vlcs}. Unless otherwise specified, each node uses a symbol period of $20$~$\mu$s {and $(216,200)$ Reed-Solomon error correction code. All the experiments are carried out in an indoor office environment in the presence of artificial lighting.

\subsection{MAC layer}
We evaluate the saturation throughput of OpenVLC's MAC layer {in} a two-node scenario, where the two nodes are within each other's Field-Of-View (FOV) and one continuously transmits to the other. The throughput as a function of the per-frame payload is shown in Fig.~\ref{fig_mac_throughput}, where the two nodes are placed at a distance of $0.6$~m and {the payload ranges from $50$ to $1000$~bytes. We measured a saturation throughput of up to $18$~kb/s. The throughput increases as the payload increases, ranging from $6$~kb/s when the payload is $50$~bytes to $18$~kb/s when the payload is increased to $1000$~bytes, which also shows the reliability of software synchronization implementation as frame size gets longer}. 

\subsection{{System Level}}
It is possible to evaluate the performance of OpenVLC using various traditional network measurement tools. In this subsection, we present evaluation results obtained from the well-known network tools \texttt{ping} and \texttt{iperf} {in point-to-point link and three-node scenarios}. 

The OpenVLC's \texttt{ping} performance {in a point-to-point link scenario} over various \texttt{ping} {Inter-Packet Interval (IPI)} settings is shown in Fig.~\ref{fig_app} (top-left). These results are collected from {$1000$} \texttt{ping} packets where each {\tt ping} packet has 10-byte data. From the empirical Cumulative Distribution Function (CDF) of the Round-Trip Time (RTT)), {we observe that when the IPI is set to $0.25$~s, about $90\%$ of the packets incur a RTT below $200$~ms. This value drops to $60\%$ when the \texttt{ping} traffic load increases to IPI=$0.2$~s as a result of the longer queuing time. }

\begin{figure*}[t!]
  \centering 
  \includegraphics[width=1.6\columnwidth]{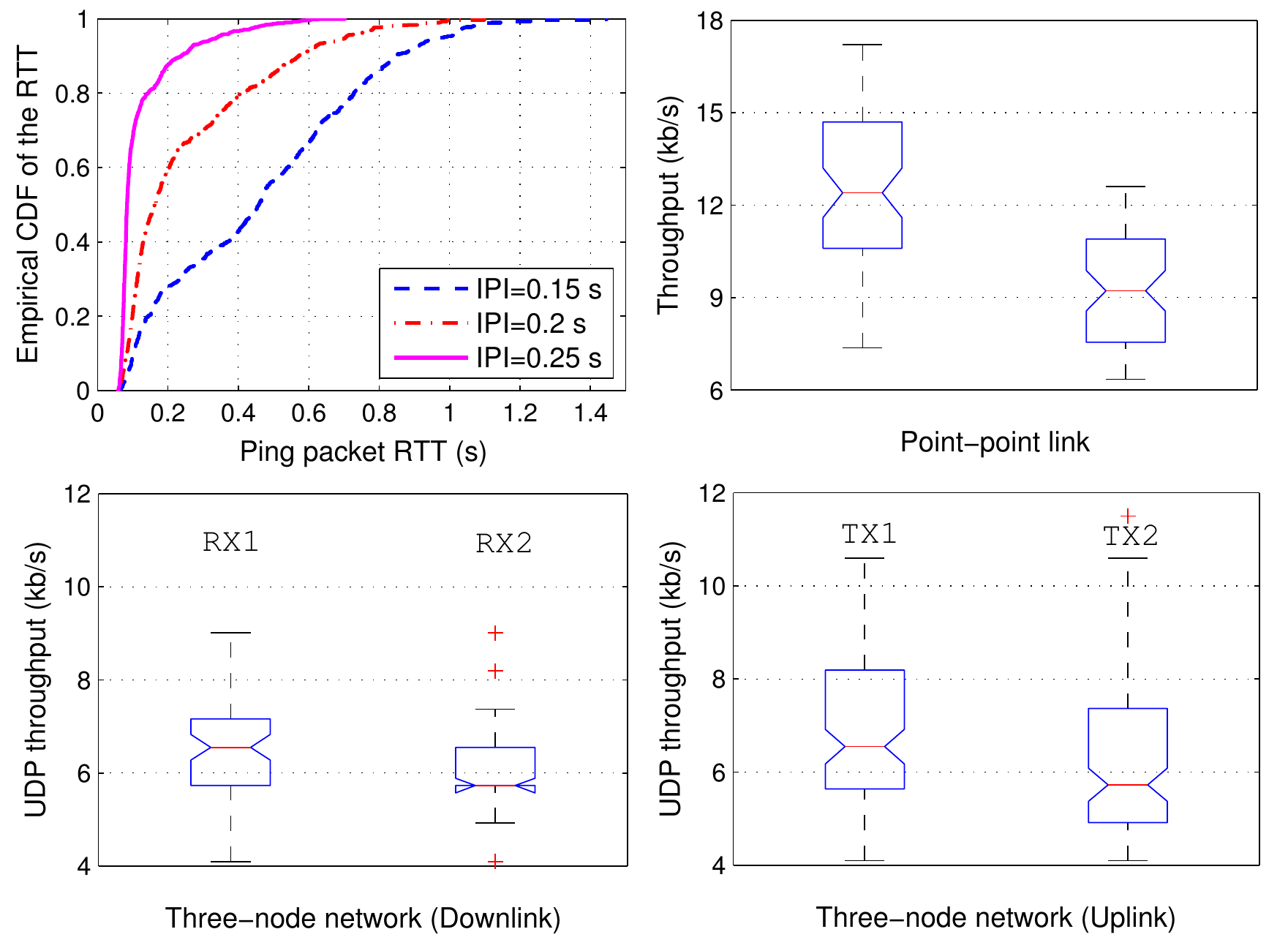}
  \caption{System-level evaluation results using \texttt{ping} and \texttt{iperf}.}
  \label{fig_app}
\end{figure*}

The network testing tool \texttt{iperf} is also used to evaluate the performance of OpenVLC, and the evaluation results of UDP and TCP {over a point-to-point link} are shown in Fig.~\ref{fig_app} (top-right). Here the UDP datagram size is set to $1000$~bytes. 
{The results are collected by running the experiment for $10$ minutes in each scenario and we plot the results reported by \texttt{iperf} every 10 seconds. We observe that the maximal and median achievable throughputs {with} UDP are about $17$~kb/s and $12.5$~kb/s, respectively. As for TCP, the maximal and median achievable throughputs are around $13$~kb/s and $9$~kb/s, respectively. The throughput drops {with} TCP with respect to UDP is due to the overhead and reliability features of TCP.}

{Furthermore, we evaluate the performance of OpenVLC in a three-node (point-to-multipoint) network, and the results are shown in Fig.~\ref{fig_app} (bottom). In the downlink scenario, one node sends UDP data to the other two nodes (RX1 and RX2). We observe that the median values of the UDP throughput of RX1 and RX2 are around $6$~kb/s. In the uplink scenario, two nodes (TX1 and TX2) compete for the shared medium to send data to the third one. The median values of the achieved throughput are also around $6$~kb/s, which shows a fair access to the medium.}

\section{Performance Enhancements and Future Research Directions of OpenVLC} \label{sec_discussion}

Currently, OpenVLC is designed using basic commercial off-the-shelf components to implement communication network among LEDs. The achieved data rate is already comparable to the lowest one specified in the IEEE 802.15.7 standard{, which specifies a PHY rate of at least $11.67$\,kb/s}~\cite{IEEE802.15.7}. In its present form, OpenVLC already offers a flexible starter kit for VLC research. 

While most of the VLC efforts so far have targeted point-to-point systems between resource-rich high-end nodes, to date, resource-poor low-end nodes are instead an unexplored research area. Exploring networked systems of resource-poor low-end nodes would be instrumental to the adoption of networked VLC and would require a fundamental redesign of the communication stack. The performance of OpenVLC can be improved to reach out other domains of research investigations, using more powerful hardware and by customizing the software implementation to the application scenarios of choice. In this section, we discuss a set of possible enhancements for OpenVLC as well as future research directions based on it.

\subsection{Performance Enhancements of OpenVLC}
We begin by reviewing a list of points that could be implemented to boost OpenVLC's performance. 

\textit{Matched filtering and timing error recovery} have not yet been implemented in OpenVLC. A matched filter serves to maximize the signal-to-noise ratio and minimize the symbol error probability. Timing error recovery is very useful when the transmitter and the receiver are unsynchronized. To support matched filtering, the front-end transceiver hardware needs to be upgraded. To implement timing error recovery, the software-defined PHY layer of OpenVLC needs to be enhanced to detect the timing error and recover from it. The implementation of the matched filtering as well as the timing error recovery on OpenVLC would also help to increase the communication range as well as the overall system stability for higher rate communication.

The \textit{coverage} of an OpenVLC node is currently limited by the output power and FOV of its LEDs. The output power can be increased {by} using high brightness white LEDs as optical front-end. For scenarios where one OpenVLC node acts as an access point, hardware should be extended to support Multiple Input Multiple Output (MIMO) LED communication, with modulations such as optical GSSK~\cite{Popoola}. This direction exploits the fact that multiple LEDs are usually required for illumination due to the limited brightness of an individual LED. The software would also need to support the selection and use of different LED-to-LED links. 

The current prototype adopts the basic OOK modulation, but \textit{advanced modulation schemes} can also be used by adding a Digital-to-Analog Converter (DAC) or by exploiting the Pulse-Width Modulation (PWM) pins of the BBB.  In this way, the disadvantage of OOK in terms of inefficient bandwidth usage can be circumvented.
For OpenVLC, the current bottleneck of the achievable date rate is the speed at which the BBB reads symbols from the ADC (the Linux system we employ fails to provide accurate timing past a certain speed~\cite{Brown2010}). In turn, the BBB can write symbols to the LED at a much faster speed. To eliminate the current bottleneck, \textit{Field-Programmable Gate Arrays (FPGAs)} {(as the one used in \cite{Videv2014} that can be interfaced with the BBB)} or \textit{micro-controllers (MCUs) could be employed for the PHY layer implementation.} Using such solutions, however, would increase the cost of OpenVLC. A cheaper alternative is to use the \textit{Programmable Real-time Unit (PRUs)} of the BBB for dedicated implementation of time-critical functionalities. The ARM CPU of the BBB has two PRUs and each PRU is a low-latency 32-bit micro-controller. To improve the performance of OpenVLC, the PRUs can be used to implement time-sensitive sampling. Because the PRUs can operate at 200 MHz, the performance gain from using them would be significant.

\begin{table*}[t!]
\centering
\small
\caption{Summary of the possible performance enhancements and research directions of the OpenVLC platform.} 
\label{table_extension}
~

\begin{tabular}{| l | l | c | c | c | c | c |} 

\hline
\tabincell{l}{Performance Enhancements \\ \& Future research directions} & \tabincell{c}{Benefits for the system} & \tabincell{c}{Need hardware \\changes?} & \tabincell{c}{Need software \\changes?} & Difficulty & \tabincell{c}{Has been \\implemented?} \\ 
\hline \hline
\tabincell{l}{Matched-filtering and \\timing error recovery} & \tabincell{l}{-Stability \\-Throughput\\-Communication distance} & Yes & Yes & Medium & No \\ \hline
\tabincell{l}{Communication coverage \\ (High brightness LED \& MIMO)} & \tabincell{l}{-Communication coverage}& Yes & Yes & Medium & No \\ \hline
\tabincell{l}{Advanced modulation scheme} &-Throughput & {No/Yes} & Yes & Medium & No \\ \hline
\tabincell{l}{FPGA for the PHY} & -Throughput& Yes & Yes & Hard & No \\ \hline
\tabincell{l}{MCU for the PHY} & -Throughput& Yes & Yes & Medium & No \\ \hline
\tabincell{l}{PRUs for the PHY (kernel space)} &-Throughput& No & Yes & Hard & No \\ \hline
\tabincell{l}{PRUs for the PHY (user space)} & -Throughput& No & Yes & Medium & No \\ \hline
\tabincell{l}{LED-to-photodiode communication} & \tabincell{l}{-To be verified} & Yes & Yes & Easy & No \\ \hline
\tabincell{l}{OpenVLC as an app} & \tabincell{l}{-Fast testing and deployment}& No & Yes & Hard & No \\ \hline
\tabincell{l}{Intra-frame bidirectional transmissions} &-Throughput & No & Yes & Medium & Yes~\cite{Wang2014conext}\\ \hline
\tabincell{l}{Integration with \\RF communication} &\tabincell{l}{-Stability\\ -Communication flexibility}& Yes & Yes & Hard & No\\
\hline
\end{tabular}
\end{table*}

\subsection{Future Research Directions Based on OpenVLC}
We will now discuss a number of promising research directions that can be pursued based on the OpenVLC.

\begin{itemize}
  \item {\textit{LED-to-Photodiode communication:} extending OpenVLC to support LED-to-Photodiode communication is straightforward. It would be very valuable to compare the performance of LED-to-LED and LED-to-Photodiode communications, in terms of transmission range, maximal achievable throughput, resilience ability to ambient light, etc.}
  
  \item \textit{OpenVLC as an app:} recent research has explored the feasibility of implementing the PHY and MAC layers of ZigBee and WiFi as downloadable pieces of software (such as apps for smartphones)~\cite{Park2014}. This approach would streamline the testing and deployment of modifications to existing protocols and, in principle, new protocols as well. With the PRUs of the BBB, it is possible to develop a software on the MAC/PHY protocols of OpenVLC within the user space of Linux without sacrificing the achievable data rate. 
 
  \item \textit{Enabling intra-frame bidirectional transmissions:} a basic choice for the PHY layer of a VLC system is the OOK modulation with the Manchester Run-Length Limited (RLL) line code. RLL line codes are used to prevent flickering. With the OOK modulation and RLL line codes, a transmitter normally does not need to emit light when it transmits a LOW symbol. As presented in Sec.~\ref{sec_mac}, the transmitter can then switch the LED to RX mode to receive a symbol. Furthermore, if the receiver has detected a HIGH symbol in current symbol slot and the HIGH symbol is the first part of a modulated bit, then the receiver can switch the LED to TX mode to transmit a symbol during the next symbol slot. Therefore, the receiver can \emph{embed data} into the current frame it is receiving. This technique enables intra-frame bidirectional transmissions that can increase the system throughput to a significant extent. We have successfully implemented this technique using OpenVLC and the details can be found in~\cite{Wang2014conext}. 

  \item \textit{Integration with RF communication:} In order to provide backward compatibility with previous embedded systems, one may think of designing hybrid communication networks that are built on top of both visible light and RF communication. This may allow to exploit the advantage of both technologies, and use them in the most appropriate channel and network conditions. 
 \end{itemize}

A summary of these research directions together with the performance enhancements of OpenVLC is given in Table~\ref{table_extension}.

\section{Conclusion} \label{sec_conclusion}

In this paper, we presented the design, implementation, and performance evaluation of OpenVLC, an open source platform designed to enable VLC research in the field of networked embedded systems. OpenVLC's paramount goal is to demystify VLC and lower the barriers to entry to VLC research for embedded systems researchers. Much like the Berkeley motes demystified low-power wireless a decade ago and paved the way to a decade's worth of rich and active research in wireless sensor networks, we believe that an open reference platform may open up the unexplored area of networked VLC for embedded devices. OpenVLC leverages the recent diffusion of powerful but cost-effective embedded Linux platforms to provide a reference platform that can be used jointly with a vast array of Linux tools. OpenVLC also shows how a handful of commercial off-the-shelf components can suffice as a starter kit for VLC research. Going forward, we hope that OpenVLC can serve as a bridge between the VLC community and the wireless embedded systems community. We envision that research groups in embedded systems with no prior VLC experience can use OpenVLC to explore the realm of visible light, while research groups with a solid VLC background can easily expand OpenVLC and enrich its set of functionalities, for instance with more sophisticated hardware and more advanced PHY designs.

 \section*{Acknowledgement}
This article has been partially supported by the Madrid Regional Government through the TIGRE5-CM program (S2013/ICE-2919).

\balance
\bibliographystyle{IEEEtran}


\section*{Biographies}

\vspace{-1.2cm}
\begin{IEEEbiographynophoto}{Qing Wang} (qing.wang@imdea.org)
is currently a PhD student with the IMDEA Networks Institute as well as the University Carlos III of Madrid. He received his Bachelor's and Master's degrees from the University of Electronic Science and Technology of China (UESTC), Chengdu, China, in 2008 and 2011, respectively. He also received a Master's degree from the University Carlos III of Madrid in 2012. His interests include visible light communication, device-to-device communication, stochastic optimization, and performance evaluation.
\end{IEEEbiographynophoto}

\vspace{-1.1cm}
\begin{IEEEbiographynophoto} {Domenico Giustiniano} (domenico.giustiniano@imdea.org)
is a Research Assistant Professor at IMDEA Networks Institute. He was formerly a Senior Researcher and Lecturer at ETH Zurich and a Post-Doctoral Researcher at Disney Research Zurich and at Telefonica Research Barcelona.  In 2008, he was awarded a Ph.D. in Telecommunication Engineering from the University of Rome Tor Vergata. Dr. Giustiniano devotes most of his current research to emerging areas in the field of wireless networking and pervasive wireless systems, including visible light communication networks and mobile indoor localization.
\end{IEEEbiographynophoto}

\vspace{-1.1cm}
\begin{IEEEbiographynophoto}{Daniele Puccinelli} (daniele.puccinelli@supsi.ch) is a Senior Research Scientist at the University of Applied Sciences and Arts of Southern Switzerland (SUPSI). He holds a Ph.D. in Electrical Engineering from the University of Notre Dame (USA). His research interests include networked embedded systems, low-power wireless networking, pervasive computing, and information technology for energy efficiency.
\end{IEEEbiographynophoto}

\end{document}